\documentclass[aps, prb, reprint, superscriptaddress, amsmath, amssymb, floatfix, twocolumn]{revtex4-2}

\let\textcite\relax
\makeatletter
\DeclareRobustCommand{\MakeUppercase}[1]{{%
      \def\i{I}\def\j{J}%
      \def\reserved@a##1##2{\let##1##2\reserved@a}%
      \expandafter\reserved@a\@uclclist\reserved@b{\reserved@b\@gobble}%
      \protected@edef\reserved@a{\uppercase{#1}}%
      \reserved@a
   }}
\DeclareRobustCommand{\MakeLowercase}[1]{{%
      \def\reserved@a##1##2{\let##2##1\reserved@a}%
      \expandafter\reserved@a\@uclclist\reserved@b{\reserved@b\@gobble}%
      \protected@edef\reserved@a{\lowercase{#1}}%
      \reserved@a
   }}
\makeatother
\expandafter\let\csname ver@natbib.sty\endcsname\relax

\usepackage{xcolor}
\usepackage{graphicx}
\usepackage{dcolumn}
\usepackage{bm}
\usepackage[section]{placeins}
\usepackage{ulem}
\usepackage{algorithm}
\usepackage{algorithmicx}
\usepackage{amsmath}
\usepackage[noend]{algpseudocode}

\usepackage[toc,page,header]{appendix}
\usepackage{minitoc}

\usepackage{tocloft}
\usepackage{setspace}
\DeclareUnicodeCharacter{2212}{-}
\usepackage[utf8]{inputenc}

\usepackage{booktabs}
\usepackage{multirow}
\usepackage{array}

\usepackage{makecell}

\begin{document}

\title{Emergent properties of van der Waals bilayers revealed by computational stacking}

\author{Sahar Pakdel, Asbjørn Rasmussen, Alireza Taghizadeh, Mads Kruse, Thomas Olsen, Kristian S. Thygesen}
\affiliation{CAMD, Computational Atomic-Scale Materials Design, Department of Physics, Technical University of Denmark, 2800 Kgs. Lyngby, Denmark.}%
\date{\today}

\begin{abstract}
    Stacking of two-dimensional (2D) materials has emerged as a facile strategy for realising exotic quantum states of matter and engineering electronic properties. Yet, developments beyond the proof-of-principle level are impeded by the vast size of the configuration space defined by layer combinations and stacking orders. Here we employ a density functional theory (DFT) workflow to calculate interlayer binding energies of 8451 homobilayers created by stacking 1052 different monolayers in various configurations. Analysis of the stacking orders in 247 experimentally known van der Waals crystals is used to validate the workflow and determine the criteria for realizable bilayers. For the 2586 most stable bilayer systems, we calculate a range of electronic, magnetic, and vibrational properties, and explore general trends and anomalies. We identify an abundance of bistable bilayers with stacking order-dependent magnetic or electrical polarisation states making them candidates for slidetronics applications. 
\end{abstract}
\maketitle

\twocolumngrid

The field of 2D materials has evolved with tremendous pace over the past decade and is currently impacting many areas of contemporary physics including spintronics\cite{yang2022two,sierra2021van}, valleytronics\cite{schaibley2016valleytronics}, polaritonics\cite{low2017polaritons}, unconventional superconductivity\cite{cao2018unconventional}, multiferroics\cite{behera2021recent}, and quantum light sources\cite{tran2016quantum}. 
While numerous 2D monolayers have been extensively scrutinised in experiments and their properties systematically organised in computational databases\cite{haastrup2018computational,mounet2018two}, studies of 2D multilayer structures have been much more sporadic. 

The unit cell commensurate homobilayers (from hereon referred to as homobilayers or simply bilayers) represent a well defined and increasingly popular class of 2D multilayer materials. Despite sharing the same Bravais lattice as the monolayer, their point group symmetry can differ depending on the stacking order. Such qualitative differences can influence physical properties profoundly with direct consequences for the material's utilization potential\cite{Liu2020,Sun2019,Cenker2021}. For example, non-volatile ferroelectric memories\cite{wang2021two}, the valley Hall effect\cite{wu2019intrinsic}, and spontaneous valley polarisation\cite{zhang20222d}, require materials with broken inversion symmetry. This may be achieved in a homobilayer even if the monolayer is centrosymmetric. Furthermore, due to the presence of the van der Waals (vdW) gap, the properties of bilayers can be tuned more effectively as exemplified by the giant Stark effect of interlayer excitons in bilayer MoS$_2$\cite{leisgang2020giant,peimyoo2021electrical}, switching of magnetic states in bilayer CrI$_3$ by either electrical gating\cite{Jiang2018,Huang2018a} or pressure\cite{song2019switching}, and coupled ferroelectricity-superconductivity in blayer MoTe$_2$\cite{jindal2023coupled}.

It was recently proposed that the layer degree of freedom in vdW bilayers could form the basis of a new type of 2D ferroelectrics\cite{li2017binary,yang2018origin,wang2023towards}. Subsequently interfacial ferroelectricity has been demonstrated in bilayers of hexagonal boron nitride\cite{vizner2021interfacial,yasuda2021stacking} and transition metal dichalcogenides (TMDs)\cite{wang2022interfacial,wan2022room,fei2018ferroelectric}. In these experiments, two stacking configurations with different out-of-plane polarisations are electrically switched via in-plane sliding of the layers. Beyond out-of-plane ferroelectricity, one might envision slide-induced switching of other physical quantities such as in-plane polarisation, conductivity, magnetism, or band topology\cite{peng2020stacking,xu2022van}. Further progress in the emerging field of slidetronics\cite{zhou2022photo,xiao2022non,zhang2022domino} calls for identification of specific bilayer systems with (meta)stable stacking configurations separated by low switching barriers.

\begin{figure*}
\centering
\includegraphics[width=0.9\linewidth]{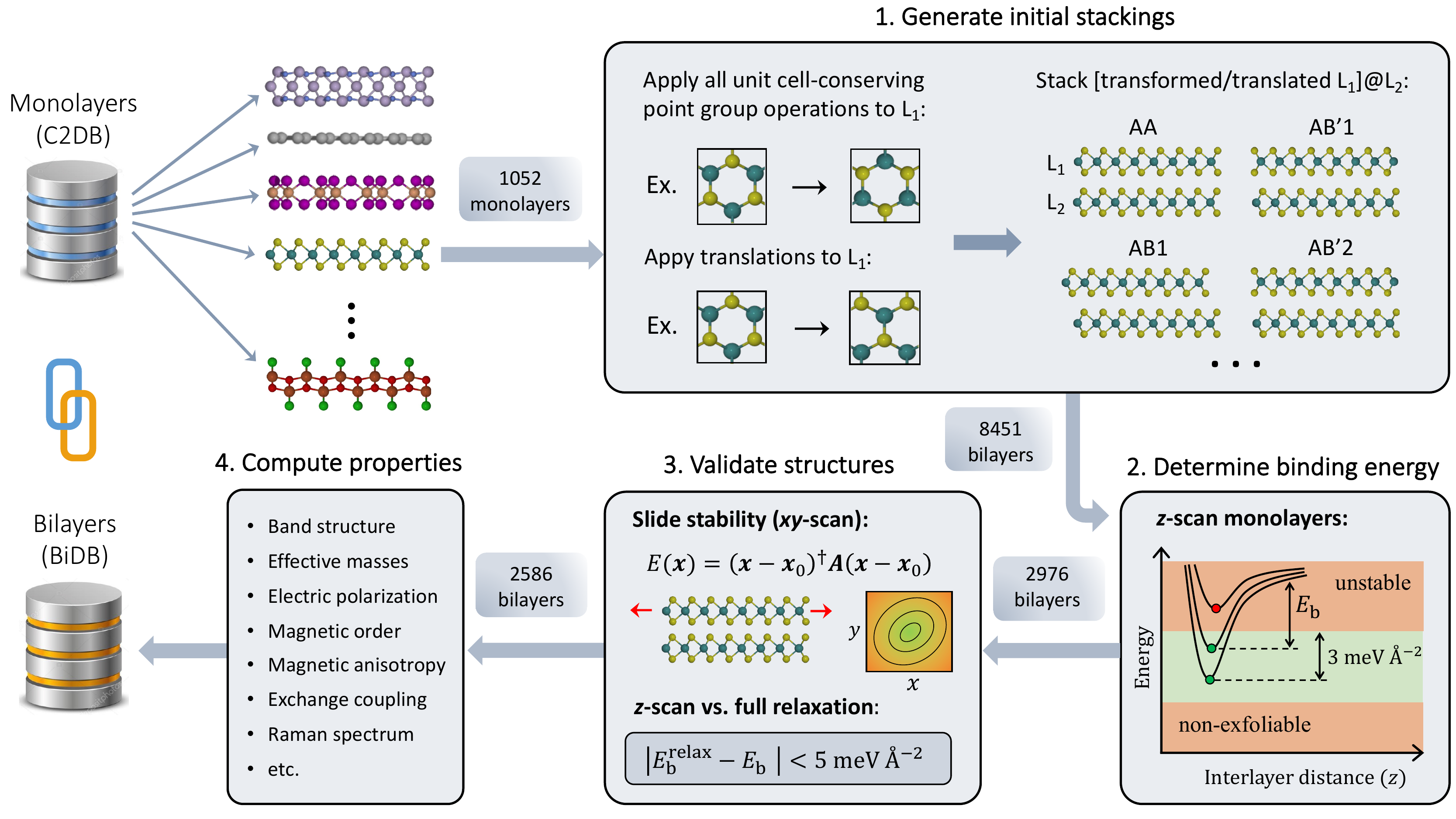}
\caption{\textbf{Stacking workflow.} A total of 1052 stable monolayers with up to 10 atoms in the unit cell are imported from the C2DB database. Starting from the AA stacked bilayer, various stacking configurations are generated by applying all unit cell-preserving point group operations to layer 1 while keeping layer 2 fixed. For each of these configurations, $N_{\mathrm{atoms}}^2$ structures are generated by translating layer 1 by a vector $\mathbf t_{ij}^{\parallel}$ given by the difference between the lateral positions of atoms $i$ and $j$ in the monolayer unit cell. Duplicate bilayers are subsequently removed. The interlayer binding energy, $E_{\mathrm{b}}$, of the 8451 unique bilayers is then computed by the $z$-scan approach. Bilayers with $E_{\mathrm{b}}$ within 3 meV/\AA$^2$ of the most stable stacking are considered thermodynamically stable. The 2976 thermodynamically stable bilayers are then validated by checking their stability against lateral sliding and the reliability of the $z$-scan approach. The final 2586 bilayers are run through the property workflow.}
\label{fig:workflow}
\end{figure*}

While the production of vdW heterostructures and twisted bilayers is a complex and time consuming process with low yield and high sample-to-sample variability, consistent high quality homobilayers may be exfoliated from naturally occurring bulk samples or grown bottom-up by scalable chemical methods\cite{shinde2018stacking,bertoldo2021intrinsic}. Despite of this clear advantage and the exciting prospects described above, the homobilayers remain largely unexplored compared to their simpler monolayer constituents. 

Here we employ a first principles high-throughput workflow to systematically construct and explore all homobilayers that can be formed from 1052 monolayers. Our bottom-up stacking workflow yields a total of 2586 bilayers that we predict to be stable and experimentally realizable based on an extensive analysis invoking the stacking orders found in 247 known vdW bulk compounds. We analyse the emergent properties of the bilayers (relative to their monolayers) and discover an abundance of systems possessing stacking order-dependent electronic, magnetic, or ferroelectric properties - some of which are switchable via layer sliding. 

All calculated crystal structures and properties are provided in an open curated database. The full library of homobilayers with all calculated properties will be provided as an open, curated database fully integrated with the C2DB monolayer database, making a unique digital platform to support and accelerate 2D materials science. \\   
%
\begin{figure*}
\centering
\includegraphics[width=0.98\linewidth]{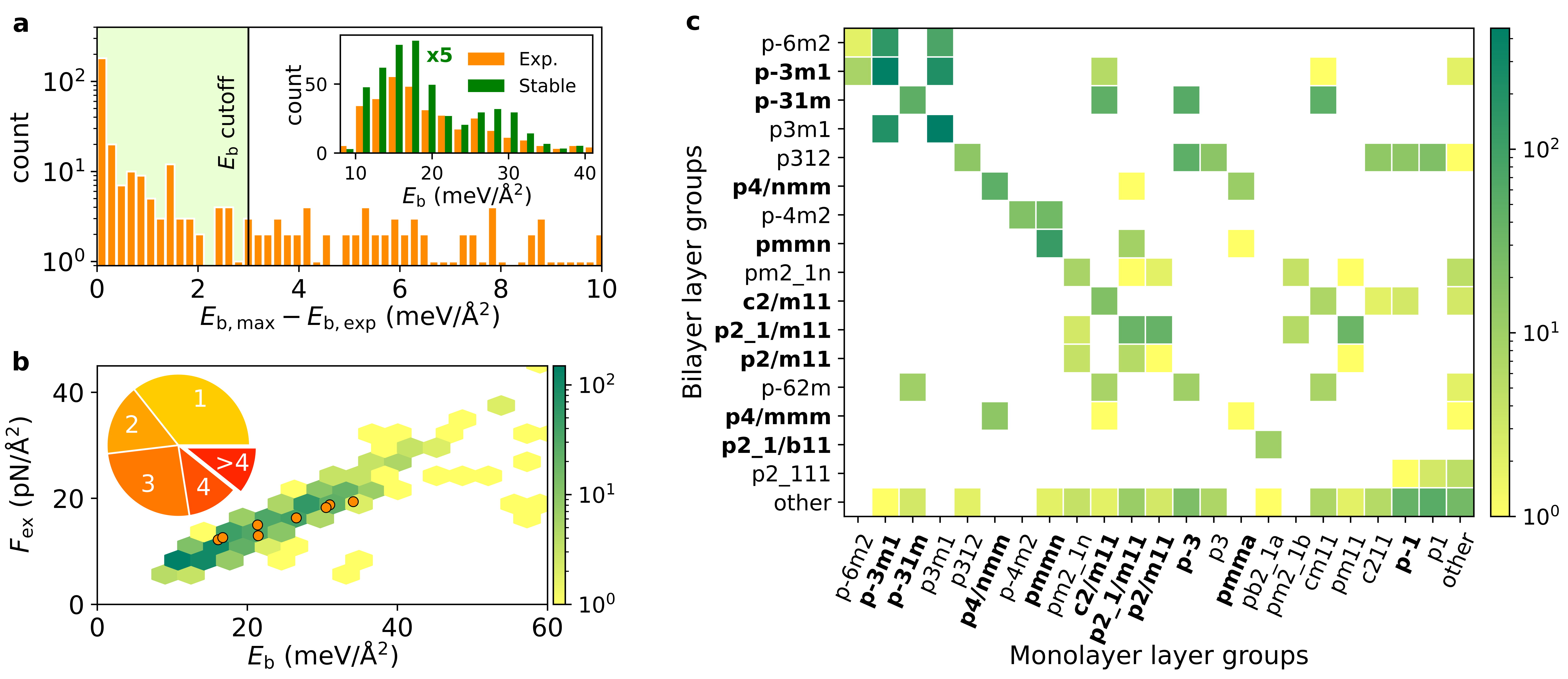}
\caption{\textbf{Bilayer stability, exfoliation force, and crystal symmetry.} (a) Histogram showing the difference between the interlayer binding energy of bilayers in an experimentally observed stacking configuration and the most stable stacking configuration predicted by the stacking workflow (note the logarithmic scale). The inset shows the distribution of the binding energies calculated for the experimentally observed stackings (orange) and all the predicted stable bilayers (green). (b) Exfoliation force versus interlayer binding energy for all stable bilayers. Forces are demonstrated for 20\% of stable bilayers which we specifically calculated more data points for the $E_{\mathrm{b}}(z)$ curve. The exfoliation force is the minimum force per area required to pull the bilayer apart. A number of known exfoliable 2D materials are indicated by orange dots (Graphene, Phosphorene, BN, MoS$_2$, NbSe$_2$, PtSe$_2$, WTe$_2$, HfSe$_2$). The inset shows the number of stable bilayers obtained per monolayer. (c) Matrix representation of the layer group of stable bilayers versus the layer group of the constituent monolayer. The layer groups with inversion symmetry are marked in bold font.}
\label{fig:structures}
\end{figure*}
%

\textbf{Stacking workflow.}
Our stacking workflow starts by extracting 1052 stable monolayers with up to 10 atoms/unit cell from the C2DB\cite{haastrup2018computational,gjerding2021recent}, and arranging them in a total of 8451 unique stacking configurations (see Fig. \ref{fig:workflow} and Methods). The interlayer distance and binding energy, $E_\mathrm{b}$, is determined for each bilayer by scanning the total energy as a function of the interlayer separation ("z-scan" approach). The vast majority of the constructed bilayers have $E_\mathrm{b}$ below 50 meV/\AA$^2$ indicating interlayer bonds of purely vdW character. This is a result of the stringent stability criteria we used to select the monolayers. A bilayer is considered thermodynamically stable if $\Delta E_\mathrm{b,max} \equiv E_{\mathrm{b,max}}-E_\mathrm{b}< 3$ meV/\AA$^2$, where $E_{\mathrm{b,max}}$ is the maximum binding energy among all the considered stacking orders. We shall return to this criterion below. Further details, motivation, and justification for the z-scan approach, including comparison to the fully relaxed structures, can be found in Methods and the SM. 

Next, the slide stability is checked for all thermodynamically stable bilayers by fitting the local potential energy surface to a second-order polynomial (see SM for details). Unstable configurations are pushed along a negative gradient (or curvature for saddle points), and a constrained energy minimisation is performed keeping the monolayers frozen. The procedure is repeated until a slide stable configuration is obtained. After removing duplicate structures, we obtain 2586 unique thermodynamically and slide stable bilayers, from hereon referred to as \emph{stable bilayers}. 

We now address the question of when a stacking configuration can be expected to be experimentally realised. Clearly, the energy of such structures should not be too high above the most stable configuration, $E_\mathrm{b,max}$. To determine a reasonable bound on $\Delta E_\mathrm{b,max}$, we analyze its value for the stacking orders observed in naturally occurring vdW crystals. Fig. \ref{fig:structures}a shows the distribution of $E_{\mathrm{b,max}}-E_{\mathrm{b,exp}}$, i.e. the binding energy of bilayers in the experimentally observed stacking relative to the most stable stacking found by our workflow (note the logarithmic scale). The analysis includes 247 experimental bulk structures from which we extract 314 bilayers (crystals with more than 2 layers per unit cell can result in several unique bilayers).  

For 73\% of the 226 monolayers, the most stable bilayer coincides with an experimentally observed stacking (leftmost bar in Fig. \ref{fig:structures}). However, there are also cases where an experimentally observed stacking has an energy above the predicted most stable configuration. This could have several reasons, e.g. differences in interlayer interactions for bulk and bilayer (in the SM we show that this effect is very small), limited accuracy of the calculations, finite temperature effects not accounted for in the calculations, or the occurrence of meta-stable stacking orders in the natural bulk crystals. 
To account for such effects while avoiding to include too many unphysical structures, we set an upper bound of $\Delta E_\mathrm{b,max}=3$ meV/\AA$^2$. With this threshold we cover 74\% of the experimentally observed stackings. 
The distributions of $\Delta E_\mathrm{b}$ for all the stable bilayers and the bilayers in experimentally observed stacking configurations are compared in the inset of Fig. \ref{fig:structures}a. The similarity of the two distributions supports our protocol for selecting stable, experimentally realisable bilayers. 

The z-scan approach allows us to fit $E_{\mathrm{b}}(z)$ by an analytical Buckingham potential for the vdW interaction between two 2D planes, and thereby determine the exfoliation force, $F_{\mathrm{ex}}$ (see Methods). Fig. \ref{fig:structures}b shows, not unexpectedly, that $F_{\mathrm{ex}}$ is correlated with $E_{\mathrm{b}}$. However, there are also deviations from a linear relationship implying that $E_{\mathrm{b}}$ is not always an accurate descriptor for exfoliability. Our large portfolio of full $E_{\mathrm{b}}(z)$ profiles could form the basis for the development of more realistic exfoliability descriptors. 

The pie chart in the inset of  Fig. \ref{fig:structures}b shows the number of stable stackings found per monolayer. Interestingly, for about 2/3 of the monolayers, our workflow predicts the existence of multiple stable stacking configurations. Such bilayers could form the basis for novel switchable systems as will be discussed later.

As mentioned, the crystal symmetry is of key importance to many applications of 2D materials. In Fig. \ref{fig:structures}c we show the change in layer group from monolayer to bilayer for the most commonly observed layer groups. The breaking/emergence of inversion symmetry (indicated by bold font) is common and clearly illustrates that qualitatively new physics may arise upon stacking.\\     
%
\begin{figure*}
\centering
\includegraphics[width=0.98\linewidth]{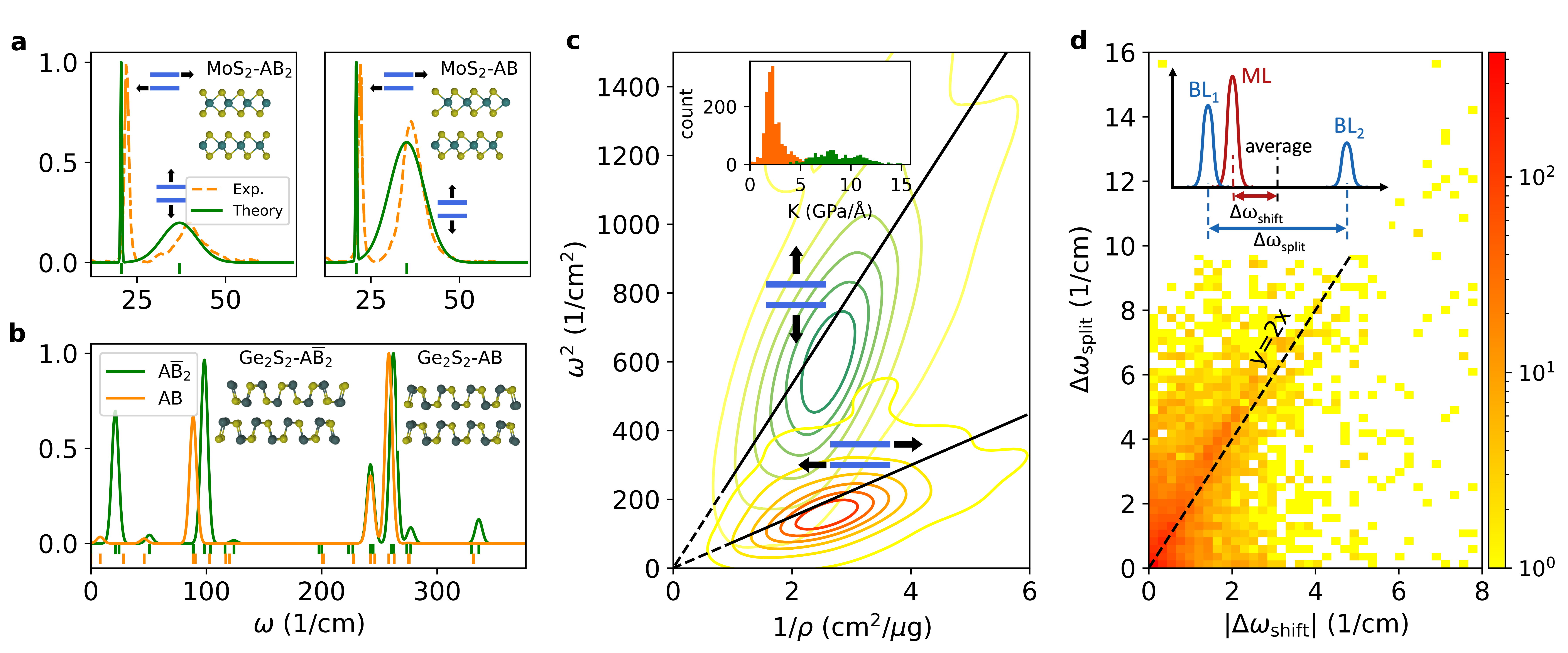}
\caption{\textbf{Raman spectra and statistics for interlayer- and intralayer vibrational modes.} (a) Low-frequency part of the Raman spectrum of bilayer MoS$_2$ in the $\mathrm A \overline{\mathrm B}_2$ and AB stacking configurations, respectively. The calculated spectra (full) are compared to the experimental spectra of Ref. \cite{van2019stacking} (dashed). The broadening of the peaks in the theoretical spectra have been fitted to match the experiments. (b) Calculated Raman spectra of bilayer Ge$_2$S$_2$ in two different stacking configurations. It can be seen that both the low and high frequency parts of the Raman spectrum is sensitive to the stacking order. (c) Distributions (contour lines) of shear and breathing interlayer modes as function of $\omega^2$ and inverse layer mass density $\rho^{-1}$. Only bilayers for which the two types of modes do not couple are included. The black lines have slopes $\langle K_\perp\rangle = 9.42$ and $\langle K_\parallel \rangle = 2.64$, corresponding to the mean value of the effective spring constants for the two distributions (shown in the inset). (d) Change in the monolayer optical phonon frequencies when stacked into bilayers. The plot shows the splitting ($\Delta \omega_{\mathrm{split}}$) versus the shift ($\Delta \omega_{\mathrm{shift}}$) of the monolayer phonon frequencies (see inset). The in-phase vibrations (BL$_1$) are essentially unperturbed while the out-of-phase vibrations (BL$_2$) are blue shifted. }
\label{fig:raman}
\end{figure*}

\textbf{Vibrational properties.}
Raman spectroscopy is one of the most important and widespread techniques for characterising 2D materials. In first-order Raman spectroscopy, the $\Gamma$-point phonons are probed via inelastic light scattering yielding detailed structural and electronic information from nondestructive measurements. While most Raman studies of layered vdW materials have focused on the high-frequency intralayer phonons, low-frequency Raman spectroscopy is emerging as a means to probe interlayer couplings with higher precision.\cite{liang2017low} To support these developments we systematically explore the sensitivity of Raman spectroscopy to interlayer coupling across the entire family of 2D materials by calculating the full Raman tensor of 481 non-magnetic monolayers and their 1244 stable bilayers. The computational methodology is described in Methods and extensive benchmarks against experiments are presented in the SM. 

Any vdW bilayer supports three low-frequency interlayer modes: Two in-plane shear modes (degenerate for isotropic materials) and one out-of-plane breathing mode. As an example, Fig. \ref{fig:raman}a shows the calculated and experimental low-frequency Raman spectrum of bilayer MoS$_2$ in $\mathrm A \overline{\mathrm B}_2$ (2H) and the AB (3R)
 stacking configurations (our bilayer notation is explained in Methods). Both the experimental and theoretical spectra show a significant difference in the relative peak amplitude in the two stacking configurations as well as a shift in frequency of the the breathing mode of 2.0 cm$^{-1}$ (calculation) and 3.5 cm$^{-1}$ (experiment). 
 Fig. \ref{fig:raman}b compares the Raman spectra of bilayer Ge$_2$S$_2$ in two stacking configurations. Again, the low frequency modes show significant frequency shifts, but even more drastic changes occur for the peak intensities due to different crystal symmetries. The high-frequency spectrum also shows differences, although to a lesser extent. These examples clearly illustrate the sensitivity of the interlayer modes to the interfacial vdW interactions, and the potential of low-frequency Raman spectroscopy for identifying stacking orders.  

Within the harmonic approximation, the frequency of an interlayer vibration fulfills 
$\omega^2=K/\rho$, where $\rho$ is the monolayer mass density and $K$ the effective spring constant quantifying the strength of the interfacial vdW interaction.
Fig. \ref{fig:raman}c shows the distributions of shear modes (red) and breathing modes (green) of all bilayers as function of $\omega^2$ and $\rho^{-1}$. Lines with slopes equal to the mean of the two $K$-distributions (see inset) are shown. The shear modes have strikingly similar spring constants while significantly larger spread is found for the breathing modes. The shown distributions include the all bilayers where the two types of interlayer modes decouple. For the remaining ca. 250 bilayers, the shear and breathing modes mix revealing an intrinsic mechanical coupling between out-of-plane uniaxial strain and interlayer shear. 

We now turn to the intralayer phonons. Each intralayer phonon of the monolayer will give rise to two phonons in the bilayer. Due to the interfacial vdW interactions, the phonons of the bilayer will split (Davydov splitting) and shift relative to the monolayer phonons, see inset of Fig. \ref{fig:raman}d. By projecting the phonon modes of the bilayer onto the modes of the isolated monolayers, we are able to track the phonon hybridisation and unambiguously determine the frequency split and shift for each individual mode. The enhanced intensity of the data points along the line $\Delta \omega_{\mathrm{split}}=2\Delta_{\mathrm{shift}}$ (note the log scale) shows that in many cases the softening of the in-phase vibrations is weaker than the hardening of the out-of-phase vibrations. The extensive data set of Raman tensors provided in this work will advance the understanding and interpretation of vibrational spectroscopy of 2D materials.   

\begin{figure*}
\centering
\includegraphics[width=1.0\linewidth]{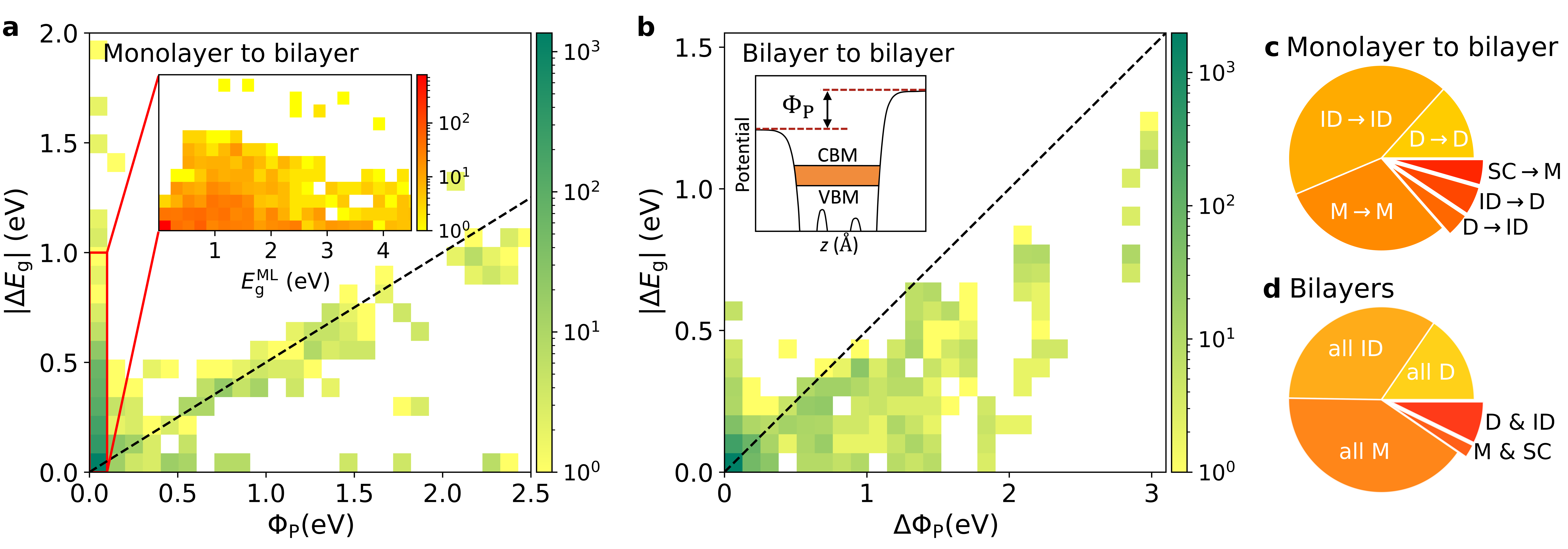}
\caption{\textbf{Trends in band gaps and electronic types.} (a) Change in electronic band gap from monolayer to bilayer as a function of the out-of-plane polarisation potential across the bilayer, $\Phi_{\mathrm{P}}$ (see inset pf panel b). Only stable bilayers are included. The dashed line indicates the situation where the bilayer gap is exactly $0.5\Phi_{\mathrm{P}}$ smaller than the monolayer gap. For the bilayers with zero out-of-plane polarisation, the gap change is shown as a function of the monolayer band gap. (b) The change in band gap between different stable stackings of the same monolayer plotted as function of the difference in the out-of-plane dipole of the two bilayers, $\Delta \Phi_{\mathrm{P}}$. (c) The distribution of bilayers according to the change in electronic type from monolayer to bilayer (M: Metal, SC: Semiconductor, D: Direct band gap, ID: Indirect band gap). There are 349 bilayers with an electronic type different from their monolayer. (d) The distribution of monolayers according to the electronic types of all its stable bilayers. There are 91 monolayers with stacking-dependent electronic types (D \& ID or M \& SC).   
}
\label{fig:gaps}
\end{figure*}

\textbf{Electronic properties.}
The idea of modifying electronic band gaps via layer stacking has been paradigmatic in the field of 2D materials. Well known examples include the indirect to direct band gap transition in the group 6 TMDs\cite{mak2010atomically} and the metal-semiconductor transition in PtSe$_2$\cite{hong2022momentum}. In general, a qualitative change in the gap type has dramatic consequences for the material's properties and possible applications. To systematically explore the opportunities and limitations of band gap stacking engineering, we have calculated the electronic band structure including spin-orbit coupling of all bilayers. Note that unstable stacking configurations are relevant as reference structures when building models for twisted bilayers. 

One mechanism affecting the band gap upon stacking is interlayer hybridisation. This effect is controlled by the overlap of wave functions across the vdW gap and as such it is sensitive to both the interlayer distance and lateral stacking configuration. In addition to interlayer hybridisation, the band gap can be affected by the presence of an out-of-plane polarisation in pyroelectric/polar bilayers. The latter effect can be quantified by the potential step across the bilayer created by the polarisation, $\Phi_{\mathrm{P}}$.  

Fig. \ref{fig:gaps}a shows the change in band gap, $\Delta E_{\mathrm{g}}$, from monolayer to bilayer as a function of $\Phi_{\mathrm{P}}$. Except for 10 bilayers, the gap is always reduced upon stacking. The common feature of the 10 anomalous monolayers is the presence of hydrogen atoms on the surface, which upon stacking interpenetrates leading to a strong and complex hybridisation. Bilayers with finite $\Phi_{\mathrm{P}}$ consist of monolayers with broken mirror symmetry (Janus monolayers) stacked with aligned dipoles. Except for the metals, characterised by $\Delta E_{\mathrm{g}}=0$, these bilayers undergo a band gap reduction of $\approx 0.5\Phi_{\mathrm{P}}$. This occurs due to the offset of the bands in the two layers produced by the polarisation potential. 
For bilayers with $\Phi_{\mathrm{P}}=0$, the gap change is solely driven by hybridisation. For this subset, the gap change shows a decreasing trend as a function of the monolayer gap (see inset). This can be understood as a reduced strength of valence-conduction band hybridisation across the vdW gap. The change in electronic type upon stacking is shown in panel (c). A total of 349 bilayers undergo either a change in band gap type or a semiconductor-metal transition upon stacking. In particular, we find 126 bilayers with an emergent direct gap (composed of indirect gap monolayers) of interest for applications in nanophotonics and opto-electronics.  

Fig. \ref{fig:gaps}b shows the difference in band gap between two stable stacking configurations of the same monolayer as function of the difference in polarisation potential, $\Delta \Phi_{\mathrm{P}}$. Despite the weakness of the vdW interaction, stacking dependent gap changes of up to 0.5 eV can arise purely from interlayer hybridisation ($\Delta \Phi_{\mathrm{P}}=0$). For bilayer pairs with finite dipole differences ($\Delta \Phi_{\mathrm{P}}>0$), gap changes of up to 1 eV occur when Janus monolayers, such as MoSSe, are stacked with parallel and anti-parallel dipole orientations, respectively. As can be seen in panel (d), most bilayers exhibit the same electronic type in all (stable) stacking configurations. However, 91 monolayers ($\sim 10\%$) support stacking configurations with different, and potentially switchable, electronic types. 

\begin{figure*}
\centering
\includegraphics[width=0.95\linewidth, trim = {0 0 0 0}]{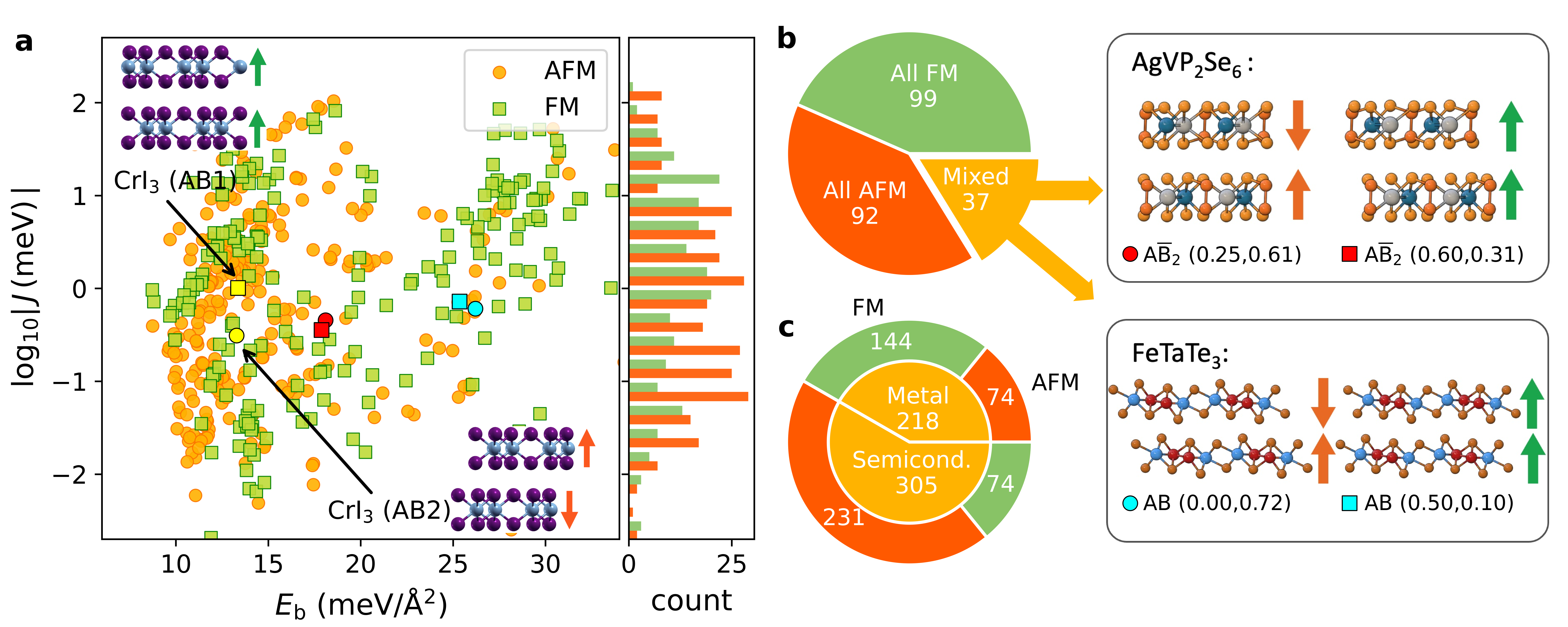}
\caption{\textbf{Magnetic properties. }(a) Interlayer exchange coupling versus interlayer binding energy for the most stable, magnetic bilayer stackings. The red (green) symbols represent bilayers with FM (AFM) order being lowest in energy. The two experimentally observed stacking configurations (also predicted as the two most stable) of bilayer CrI$_3$ are shown. (b) Fraction of magnetic monolayers for which the magnetic order of \textbf{all} stable stackings is either FM (green), AFM (orange), or both (yellow). (c) Distribution of all stable bilayer stackings according to magnetic order (FM/AFM) and electronic type (metal/semiconductor). }
\label{fig:mag2}
\end{figure*}

\textbf{Magnetic properties}
Magnetically ordered bilayers constitute a highly attractive platform for designing versatile building blocks in next-generation spintronic devices, such as spin filter magnetic tunnel junctions\cite{song2018giant} and spin tunnel field effect transistors\cite{jiang2019spin}. 
The interfacial vdW coupling allows for possible ferromagnetic (FM) or antiferromagnetic (AFM) orders. We write the energy difference between the two magnetic states as
\begin{equation}
E_{\mathrm{FM}}-E_{\mathrm{AFM}} = 2N_\mathrm{a}JS^2,
\end{equation}
which defines the effective interlayer exchange coupling $J$. Here $S$ is the average spin per magnetic atom and $N_\mathrm{a}$ is the number of magnetic atoms in the monolayer unit cell, see SM for details. Accurate determination of $J$ from first principles is challenging as it depends sensitively on the stacking configuration, temperature, and the model used for exchange-correlation effects. In the SM we show that our calculations correctly predict the experimentally observed magnetic order (FM or AFM) in seven out of nine magnetic homobilayers.

Figure \ref{fig:mag2}a shows $J$ versus $E_{\mathrm{b}}$ for 595 stable magnetic bilayers. The lack of any clear correlation between the two quantities suggests that they are governed by different physical mechanisms. Indeed, $E_{\mathrm{b}}$ is governed by vdW interactions, which is a non-local correlation effect, while $J$ is an exchange effect, which depends on the spatial overlap of the orbitals carrying the magnetic moments in the two layers. 

We find 35 monolayers with stable bilayers exhibiting FM or AFM order depending on the stacking configurations (Figure \ref{fig:mag2}b). One of these materials is the well known CrI$_3$. Among the remaining bilayers we highlight AgVP$_2$Se$_6$ and FeTaSe$_3$. Both materials are experimentally known in bulk form and the relatively small $E_{\mathrm{b}}$ indicate that they should be exfoliable. According to the C2DB, monolayer FeTaTe$_3$ is metallic with in-plane magnetic easy axis, while monolayer AgVP$_2$Se$_6$ is a semiconductor with FM order, out-of plane easy axis and in-plane ferroelectric order\cite{kruse2022two}. For both materials, the interlayer FM and AFM stacking configurations are related by a pure translation implying that the magnetic state could be switched by sliding.   

Interestingly, monolayers with broken mirror symmetry also form bilayers with stacking-dependent magnetic order. For example, bilayers of the Janus monolayer VTeS are metallic and the built-in dipole leads to charge transfer, which affects the magnetic properties of the two constituent monolayers. In particular, one stacking yields a ferrimagnetic bilayer due to the changes in the magnetic moments induced by the interlayer coupling. In general, magnetic Janus heterostructures are expected to comprise rich possibilities for controlling intrinsic magnetic properties.\\
Figure \ref{fig:mag2}c shows the distribution of all the stable magnetic bilayers according to magnetic order (FM/AFM) and electronic type (metallic/semiconducting). It can be seen that among the semiconductors there is a strong tendency for AFM order while the metals have mainly FM order.\\
\textbf{Ferroelectric switching.}
Ultrathin vdW materials with bistable stacking configurations could form the basis of ferroelectric devices such as fast non-volatile memories\cite{li2017binary,yang2018origin,wang2023towards}.
Recent experiments have demonstrated interfacial ferroelectric (IF) switching in bilayers of hBN, some group-VI TMDs and few more materials (see Ref. \cite{wang2023towards}). Beyond those systems IF switching in nine homobilayers of AB honeycomb structures were previously studied using first-principles calculations\cite{wang2023sliding}. 

\begin{figure}
\centering
\includegraphics[width=1.0\linewidth, trim = {0 0 0 0}]{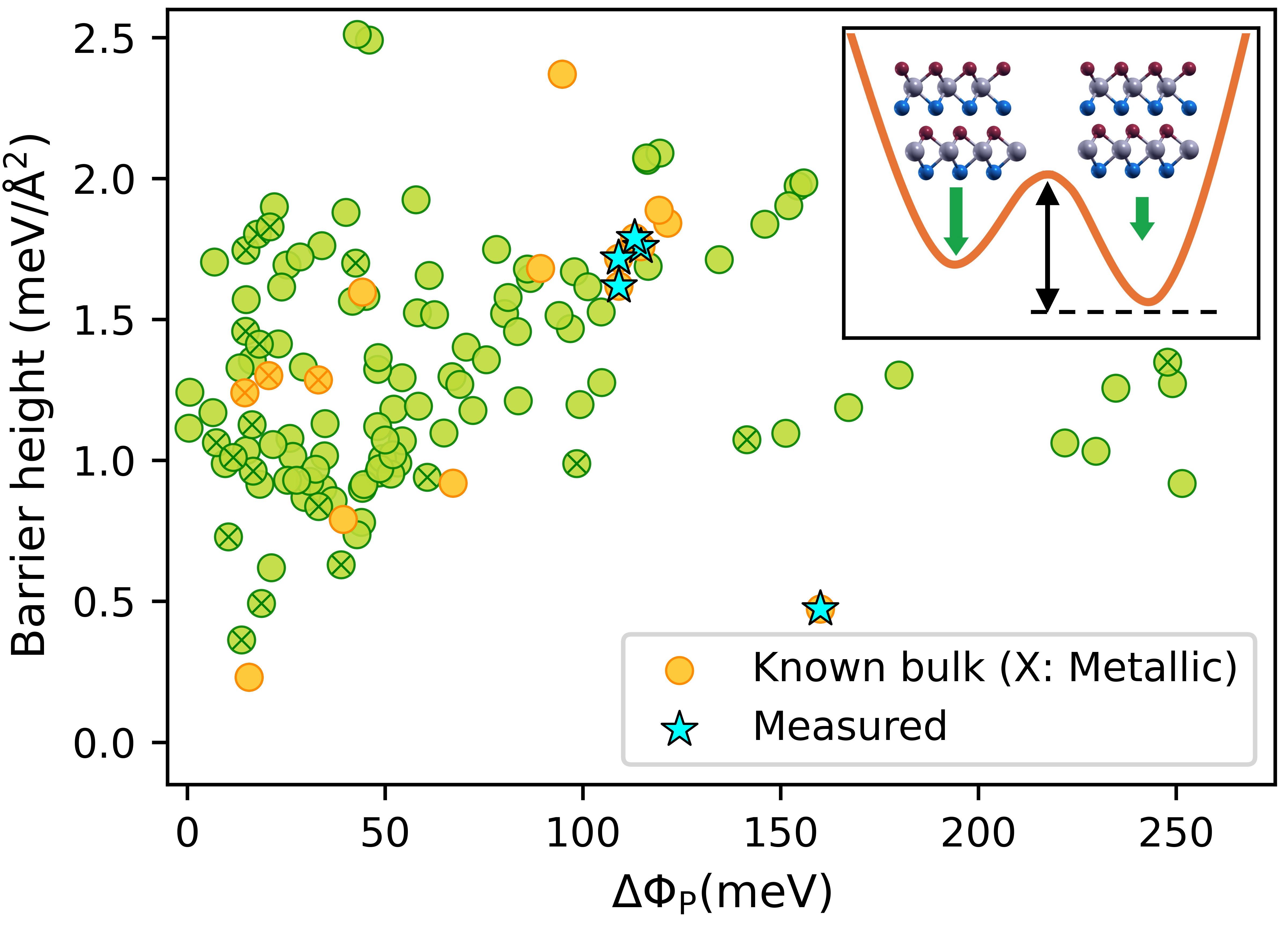}
\caption{\textbf{Interfacial ferroelectricity.} Calculated energy barrier and polarisation change between pairs of stable stacking configurations related by a pure layer translation for bilayers composed of hexagonal AB, AB$_2$ and ABC monolayers. Materials with a known bulk parent are shown in yellow. A cross on a circle indicates that the bilayers are metallic. Materials for which ferroelectric switching has been experimentally demonstrated (hBN, MoS$_2$, MoSe$_2$, WS$_2$, and WSe$_2$) are indicated by blue stars. Inset: Sketch of potential energy surface for a bistable bilayer. The green arrows indicate the out-of-plane polarisation in the two stacking configurations.}
\label{fig:switchable}
\end{figure}

Our database introduces over 1600 pairs of bilayers with at least two stable stacking configurations related by interlayer translation. To explore the phenomenon of IF more systematically, we have calculated the barrier height and change of out-of-plane polarisation, $\Delta \Phi_{\mathrm{P}}$, for the subset of bistable bilayers composed of hexagonal AB, AB$_2$, and ABC monolayers that support two stable and slide-equivalent stacking configurations, see Figure \ref{fig:switchable}. 
Within this class we identify 133 bilayer pairs with barriers below 3 meV/\AA$^2$. These include the experimentally known systems for which our calculated polarisation change, $\Delta \Phi_{\mathrm{P}}$, are in good agreement with the measured values (in meV): 160/218 (hBN), 114/94 (MoS$_2$), 108/114 (MoSe$_2$), 109/112 (WS$_2$), 113/112 (WSe$_2$) for calculation/experiment.

As can be seen from Figure \ref{fig:switchable}, we find several new candidates including some with experimentally known bulk phases (yellow symbols). Interestingly, we find several bilayers composed of ABC Janus monolayers, indicating that the build-in dipole of the monolayer promotes IF. For example the semiconductors, BiITe, which has been exfoliated in few-layer form\cite{fulop2018exfoliation}, is bistable with $\Delta \Phi_{\mathrm{P}}=90$ meV. Moreover, MoSSe, which has been synthesised as monolayer\cite{lu2017janus}, exhibits IF in two distinct rotational phases corresponding to twist angles of 0 and 60 degrees and with $\Delta \Phi_{\mathrm{P}}$ of 120 and 146 meV, respectively. A complete overview of all stacking configurations of MoSSe and other selected bilayers can be found in the SM benchmark section.  

\textbf{Conclusion}
We provided a comprehensive and systematic overview of 8451 unit cell commensurate vdW homobilayers created by combining 1052 monolayers in various stacking configurations. By analysing the stacking orders occurring in 247 natural vdW crystals, we established criteria for bilayers to be stable and experimentally realisable. For 2586 stable bilayers we calculated the electronic band structure, out-of-plane polarisation, interlayer magnetic exchange coupling as well as the Raman tensor for bilayers with up to 8 atoms/unit cell. The low-frequency Raman spectrum was found to be particularly sensitive to the interlayer vdW coupling and to provide a unique fingerprint of the stacking configuration. An abundance of bilayers with emergent properties was identified, e.g. 126 direct band semiconductors composed of monolayers with indirect gap. Moreover, a number of bistable bilayer systems with stacking-dependent polarisation or magnetic order were discovered and proposed as candidates for slidetronics applications. 
By virtue of their vdW gap and stacking order degree of freedom, the unit cell commensurate homobilayers 
offer novel functionalities and expand the space of 2D materials beyond the monolayer paradigm. 



\section{Methods}

\textbf{Workflow and data reproducibility.} 
The computational workflow was constructed within the Atomic Simulation Recipes (ASR) Python framework\cite{gjerding2021atomic} and executed using the MyQueue\cite{mortensen2020myqueue} job scheduler. The ASR workflow employs the GPAW\cite{enkovaara2010electronic} electronic structure code and the Atomic Simulation Environment (ASE) Python library\cite{larsen2017atomic}. All data discussed in this paper can be reproduced by running the ASR workflow.\\

\textbf{Density functional theory calculations.}
All DFT calculations were performed with the GPAW\cite{enkovaara2010electronic} code using (unless otherwise stated) a plane wave basis set with cutoff energy 800 eV, a uniform k-point grid of density 12.0/$\mathring{\rm A}^{-1}$, and a Fermi–Dirac smearing of 50 meV. All monolayers were relaxed using the Perdew–Burke–Ernzerhof (PBE) xc-functional\cite{perdew1996generalized} with spin polarisation. Interlayer distances and binding energies were obtained using the PBE-D3\cite{grimme2010consistent}. A minimum vacuum region of 15 Å was used to separate the periodically repeated bilayers.

After structure determination and stability analysis, a static ground state calculation is performed with a tight convergence criterion of 10$^{-6}$ eV/electron. The resulting density is used for the non-selfconsistent calculations of various properties. Spin-orbit coupling is included when evaluating band energies.   
In case of magnetic systems, the ground state was calculated for both the FM and AFM configurations to obtain the most stable (FM/AFM) magnetic state and the size of the interlayer magnetic exchange coupling, $J$. 

For all bilayers containing one of the transition metal atoms V, Cr, Mn, Fe, Co, Ni, or Cu, if the monolayer exhibits a finite band gap at the PBE level, a Hubbard-$U$ correction of 4.0 eV was applied to the 3d orbitals in the calculation of band structures and magnetic properties. The condition on monolayer gap is invoked because PBE+U is not justified for systems with metallic screening. In the SM we show that this approach yields results in good agreement with experiments for 
a number of known magnetic homobilayers. \\

\textbf{Generation of homobilayers.}
To generate an initial set of possible stacking configurations, we first determine the point group of the Bravais lattice of the monolayer, i.e. all transformations that leave the unit cell of the monolayer unchanged. The set of such transformations is denoted by $\mathcal S$. Next, we construct all 2D translation vectors of the form $\mathbf t_{ij}=\mathbf r^{\parallel}_i-\mathbf r^{\parallel}_j$, where $\mathbf r^{\parallel}_i$ is the in-plane position of atom $i$ in the monolayer unit cell. The set of such vectors is denoted 
$\mathcal T$. Starting from the AA stacked bilayer, we generate new bilayer structures by applying the transformations $\{\mathbf t \circ S \,\,|\, S\in \mathcal S,\, \mathbf t\in \mathcal T\}$ to the upper layer. Since $S$ may contain inversion/reflections, we use the PyMatGen structure matcher\cite{ong2013python} to verify that the transformation did not deform the monolayer. For each Bravais lattice, the set $\mathcal T$ is amended by a fixed (material independent) set of translation vectors. Duplicate structures are removed using the PyMatGen structure matcher.\\ 

\textbf{Bilayer notation.}
The bilayers are obtained by applying a transformation of the form $\mathbf t \circ S$ to the upper layer of the AA stacked bilayer (the fixed point of the point group operation is the origin). The point group operation, $S$, can always be written as a product of at most three basic operations: rotation of $2\pi/n$ around the out-of-plane axis ($R_n$), reflection in the horisontal plane of the 2D material ($\sigma_h$), and reflection in a vertical plane ($\sigma_v$). Note that inversion ($\mathbf r \to -\mathbf r$) is $i=\sigma_h R_2$ and rotation of $\pi$ around an in-plane axis is $\sigma_v\sigma_h$. In the most general case the resulting bilayer is denoted $\mathrm{A}\overline{\mathrm{B}}^v_n (\mathbf t$), where the bar denotes $\sigma_h$, the superscript $v$ denotes $\sigma_v$, and the subscript $n$ denotes $R_n$. 

Examples of notation: A pure sliding of the upper layer is denoted AB ($\mathbf t$), and the special case AB ($\mathbf 0, 0)$ is denoted AA. A reflection of the upper layer with respect to the horizontal plane
without any translation is denoted $\mathrm A \overline{\mathrm A}$ ($\mathbf 0, \mathbf 0$). The 3R and 2H stacking configurations of MoS$_2$ are denoted AB $(0.33,0.67)$ and $\mathrm A \overline{\mathrm B}_2$ $(0.67,0.33)$ respectively. Note that $\overline{\mathrm B}_2$ is equivalent to an inversion of the layer.

We note that the value of the translation vector $\mathbf t$ depends on the choice of the monolayer unit cell and on the lateral position of the monolayer within the cell. Despite of this ambiguity, the bilayers in the BiDB are uniquely defined by the transformation encoded in our notation when applied to the monolayer structure and unit cell from the C2DB.\\
\textbf{The z-scan approach.}
For each stacking configuration we optimise the interlayer distance by minimizing the total energy calculated with the PBE-D3 xc-functional\cite{grimme2010consistent} while keeping the monolayers fixed in their PBE-relaxed structure. For non-magnetic systems, a SciPy optimisation was employed to obtain the interlayer distance with a minimal number of DFT calculations. However, for magnetic systems this approach can lead to unphysical magnetic configurations. Hence, in such cases we reduce the distance between the monolayers in steps of decreasing size starting from a layer separation of 5 Å (layer separation is here defined as the minimal vertical atom-atom distance). The z-scan calculations employed a uniform k-point grid of density 6.0/$\mathring{\rm A}^{-1}$ and an energy convergence of 10$^{-4}$ eV/electron. 

As a byproduct of the z-scan approach we obtain a sampling of the binding energy as function of interlayer distance, $E_\mathrm{b}(z_n)$. To obtain the exfoliation force we fit $E_\mathrm{b}(z_n)$ by a Buckingham potential\cite{li2019two} describing the vdW interactions between two 2D planes. The exfoliation force is then obtained as the maximum of the derivative of the Bucking potential. We note that for bilayers with large binding energies (i. e. $>30$ meV/\AA$^2$), the Buckingham potential might not be a good fit to the entire binding energy curve. However, since we have more data points in the region around the minimum, the fitted curve is still reliable for estimating the exfoliation force.

The choice of the PBE-D3 xc-functional is motivated by its simplicity, consistency with the PBE used for the monolayer structures, and its relatively good performance for interlayer binding energies and distances in layered materials\cite{tran2019nonlocal}. In the SM we provide further justification for the choice of xc-functionals and the z-scan approach. \\

\textbf{Slide stability.}
The slide-stability of a bilayer is inferred from the local curvature of the 2D potential energy surface (PES). The latter is sampled by minimum 8 points obtained by sliding the top layer laterally in steps of 0.1$\mathring{\rm A}$ (in some cases step sizes of 0.05 or 0.15 $\mathring{\rm A}$ were included to obtain an accurate second order fit). For bilayers corresponding to a saddle point of the PES, a constrained relaxation (with frozen monolayers) was performed until the total force on one monolayer was below 0.01 eV/Å. The finite-difference sliding test was then repeated to ensure the stability of the new structure. The computational parameters were the same as used for the ground state calculations. More details on the slide stability workflow can be found in the SM.\\

\textbf{Full relaxation of bilayers.}
The z-scan calculations were supplemented by PBE-D3 calculations in which the bilayers were fully relaxed until the maximum force on any atom was below 0.01 $eV/\mathring{\rm A}$. The mean absolute deviation between the interlayer binding energies obtained with the two approaches is only 0.92 meV/\AA$^2$ (4.3\%). The very good agreement shows that intralayer relaxations driven by the layer-layer interactions are weak. We stress, that the fully relaxed structures are not necessarily more accurate than the z-scan structures, as the PBE-D3 may not be as accurate as the PBE for the strong in-plane bonds. However, for a given bilayer structure of interest it may be relevant to compare the structures and binding energies obtained with the two approaches to assess the effect of interlayer coupling-induced relaxations. More details can be found in the SM.\\

\textbf{Raman calculations.}
First-order Raman spectra were calculated for the non-magnetic bilayers with up to 16 atoms in the unit cell. The calculations are done in the Kramers–Heisenberg–Dirac approximation, where the Raman tensor $R_{ij}^\nu$ is obtained as the derivative of the electric susceptibility $\chi_{ij}^{(1)}$ along the $\Gamma$-point phonon modes \cite{lee1979time},
\begin{equation}
    \label{eq:raman_tensor}
    R_{ij}^\nu = \sum_{\alpha l} \frac{\partial \chi_{ij}^{(1)} }{\partial r_{\alpha l}} \frac{v_{\alpha l}^\nu}{\sqrt{M_\alpha}} \, .
\end{equation}
Here, $r_\alpha$ and $M_\alpha$ are the position and atomic mass of atom $\alpha$, respectively, and $v_{\alpha l}^\nu$ is the eigenmode of phonon $\nu$. 
The phonon modes were calculated using the the PBE-D3 xc-functional for proper description of the vdW-governed interlayer modes (using Phonopy package). The susceptibility was calculated within the random phase approximation (RPA) using a 800 eV plane wave cutoff, including conduction bands twice the number of valence bands. For phonon and susceptibility calculations k-meshes with the density of 6 and 15 \AA$^{-1}$ for ground state calculations were chosen. The electric susceptibility tensor, and Raman tensor, are calculated at various important incident wavelengths (488, 532, 780, 980, 1064, 1550) nm with a broadening of 200 meV. The Raman intensity is then calculated for input/output electromagnetic fields with polarization vectors $u_\mathrm{in/out}^i$ using \cite{taghizadeh2020library}
\begin{equation}
    \label{eq:raman_intensity}
    I(\omega) = I_0 \sum_\nu \frac{n_\nu+1}{\omega_\nu} \bigg|\sum_{ij} u_\mathrm{in}^i R_{ij}^\nu u_\mathrm{out}^j \bigg|^2 \delta(\omega-\omega_\nu) \, .
\end{equation}
where $I_0$ is an unimportant constant, and $n_\nu$ is obtained from the Bose--Einstein distribution, i.e.\ $n_\nu \equiv (\exp[\hbar\omega_{\nu}/k_BT]-1)^{-1}$ at temperature $T$ (here set to 300 K) for a Raman mode with energy $\hbar\omega_\nu$. The Raman peaks were represented by Gaussians of width 3 cm$^{-1}$ (replacing the delta function in Eq.~(\ref{eq:raman_intensity})), which accounts for the spectral broadening of the phonon modes.\\

\textbf{NEB calculations.}
In order to identify possible interfacial ferroelectric bilayers,
we start by screening the bilayers for bi-stable pairs (two meta-stable stackings related by a translation). Importantly, we also consider the possibility of a bilayer switching to an in-plane mirror image of itself (e.g. AB/BA stacking of hBN). We then confirm that the bilayer pairs remain slide-switchable after full relaxation. 
To calculate the barrier heights, we perform regular nudge elastic band (NEB)\cite{henkelman2000improved} calculations followed by a climbing NEB\cite{henkelman2000climbing} with maximum 7 images along the slide vector between the initial and final stable bilayers. 
We used BFGS and FIRE as local optimization algorithms available in ASE for regular and climbing NEB respectively.
For these calculations we employed a uniform k-point grid of density 10.0/$\mathring{\rm A}^{-1}$ and an energy convergence of 10$^{-6}$ eV/electron and we relaxed the structures until the maximum force on any atom was below 0.01 $eV/\mathring{\rm A}$. \\

\textbf{Data availability.}
All calculated crystal structures and properties (interlayer binding energies, electronic band structures, out-of-plane dipole moments, magnetic moments, interlayer exchange couplings, and Raman spectra) will be made available via an online database. The bilayer database will be fully integrated with the C2DB making it easy to explore monolayer and bilayer properties within one coherent framework. The computational methods and settings of the workflows used to generate the two databases are fully consistent, which is crucial for reliable quantitative comparisons.\\

\textbf{Acknowledgements}
The authors would like to thank Ask H. Larsen, Jens J. Mortensen, and Ole H. Nielsen for technical assistance with software codes and HPC access. We acknowledge funding from the European Research Council (ERC) under the European Union’s Horizon 2020 research and innovation program Grant No. 773122 (LIMA) and Grant agreement No. 951786 (NOMAD CoE).
K. S. T. is a Villum Investigator supported by VILLUM FONDEN (grant no. 37789).\\

\textbf{Author contributions}
Sahar Pakdel designed the final workflow, performed most of the calculations and data analysis, made all figures, and wrote the first draft of the manuscript. Asbjørn Rasmussen developed a first version of the workflow and carried out preliminary calculations. Alireza Taghizadeh performed some of the Raman calculations. Mads Kruse helped creating the benchmark tables with comparison to literature data. Thomas Olsen assisted with the code design and analysis of magnetic exchange couplings. Kristian S. Thygesen conceived and supervised the entire project.\\

\textbf{Competing interests}

The authors declare no competing interests.\\

\textbf{Additional information}

\bibliographystyle{apsrev4-2}
\bibliography{references.bib}

\end{document}